\documentclass[aps,superscriptaddress,preprint,nofootinbib]{revtex4}

\usepackage{graphicx}
\usepackage{bm}
\usepackage{amsmath,epsfig}

\begin{document}



\title{Spanning Trees and bootstrap reliability estimation in correlation based networks}

\author{Michele Tumminello}
\affiliation{Dipartimento di Fisica e Tecnologie Relative, 
Universit\`a di Palermo, Viale delle Scienze, I-90128 Palermo, Italy}

\author{Claudia Coronnello}
\affiliation{Dipartimento di Fisica e Tecnologie Relative, 
Universit\`a di Palermo, Viale delle Scienze, I-90128 Palermo, Italy}

\author{Fabrizio Lillo}
\affiliation{Dipartimento di Fisica e Tecnologie Relative, 
Universit\`a di Palermo, Viale delle Scienze, I-90128 Palermo, Italy}
\affiliation{Santa Fe Institute, 1399 Hyde Park Road, Santa Fe, NM 87501, U.S.A.}
\affiliation{Istituto Nazionale di Fisica Nucleare, Sezione di Catania, Catania, Italy}

\author{Salvatore Miccich\`e}
\affiliation{Dipartimento di Fisica e Tecnologie Relative, 
Universit\`a di Palermo, Viale delle Scienze, I-90128 Palermo, Italy}
\affiliation{Istituto Nazionale di Fisica Nucleare, Sezione di Catania, Catania, Italy}

\author{Rosario N. Mantegna}
\affiliation{Dipartimento di Fisica e Tecnologie Relative, 
Universit\`a di Palermo, Viale delle Scienze, I-90128 Palermo, Italy}
\affiliation{Istituto Nazionale di Fisica Nucleare, Sezione di Catania, Catania, Italy}

\date{\today}

\begin{abstract}
We introduce a new technique to associate a spanning tree to the average linkage cluster analysis. We term this tree as the Average Linkage Minimum Spanning Tree. We also introduce a technique to associate a value of reliability to links of correlation based graphs by using bootstrap replicas of data. Both techniques are applied to the portfolio of the 300 most capitalized stocks traded at New York Stock Exchange during the time period 2001-2003. We show that the Average Linkage Minimum Spanning Tree recognizes economic sectors and sub-sectors as communities in the network slightly better than the Minimum Spanning Tree does. We also show that the average reliability of links in the Minimum Spanning Tree is slightly greater than the average reliability of links in the Average Linkage Minimum Spanning Tree.
%
 \end{abstract}

\maketitle

\section{Introduction}

The study of multivariate data recorded by monitoring a wide class of physical \cite{Forrester94,Demasure03}, biological \cite{Holter00,Alter00,Holter01} and economic systems \cite{Laloux99,Plerou99,Mantegna99} is an interesting and challenging task in the investigation of complex systems. Many efforts have been done in the analysis of multivariate time series, several of them focusing on the study of pair cross-correlation between elements of the system. In the analysis of cross-correlation one faces with statistical uncertainty due to the finite length of time series, with the noise due to the interaction of the system with the environment and also with the intrinsic complexity of interactions among elements of the system. Attempts to overcome these problems may require to filter out statistically reliable information from the correlation matrix. Classical spectral methods \cite{Mardia} and related refinements \cite{Laloux99,Plerou99}, clustering \cite{Anderberg} and graph theory \cite{West} revealed to be fruitful approaches to the analysis of empirical data of correlation based systems \cite{Mantegna99,Tumminello05,Onnela02,Onnela03,Onnela04,Holter00,Ross00, Alizadeh00,Eisen98,Coronnello05,Giada01,Marsili02,Bonanno01,Bonanno03,Bonanno04}. In this paper we exploit the deep relation between the Single Linkage Cluster Analysis (SLCA) \cite{Anderberg} which is a classical technique of hierarchical clustering, and the Minimum Spanning Tree (MST) \cite{West} which is a classical concept of graph theory \cite{Gower69}, to introduce a new tree that we name Average Linkage Minimum Spanning Tree (ALMST). This graph is deeply related to another classical technique of hierarchical clustering, namely the Average Linkage Cluster Analysis (ALCA) \cite{Anderberg}. This method is widely used in phylogenetic analysis where it is known as ``unweighted pair group method using arithmetic averages" (UPGMA) \cite{Sokal58}. The motivation for introducing the ALMST as a counterpart in graph theory of ALCA bears both on the relevance of the ALCA in the study of correlation based systems such as gene expression data \cite{Holter00,Ross00, Alizadeh00,Eisen98} and stock returns \cite{Coronnello05} and on the extra information carried by the MST (ALMST) with respect to the SLCA (ALCA)\footnote{Indeed hierarchical clustering disregards the role of single elements in the cluster merging process while in spanning trees such a role is taken into account.}. In our investigation we have observed that the MST and ALMST are sharing information usually accounted by a significant percentage of common links. 
It is then worth to understand and measure the reliability of links accounting for both the finite length of time series and the data distribution. We propose here to measure such a reliability with the bootstrap technique. This technique is widely used in phylogenetic analysis as a measure of phylogenetic reliability but it has never been applied to correlation based graphs to our knowledge. Striking advantages of the bootstrap approach to test the link reliability are (i) that it does not need to introduce assumptions about the data distribution and (ii) that it naturally takes into account the length of time series. The only disadvantage that we can recognize is that the procedure is rather computationally time consuming for large networks. 

The paper is organized as follows. In section \ref{review} we review some of the most popular tools used to filter out information in correlation based systems. In section \ref{methods} we discuss both the construction of ALMST and the bootstrap approach we introduce to measure the link reliability.
In section \ref{empirical} we apply the tools introduced in section \ref{methods} to daily price returns of the 300 most capitalized stocks traded at New York Stock Exchange (NYSE) during the time period 2001-2003. 
Finally in section \ref{conclusions} we draw our conclusions.

\section{Filtering information from correlation matrices}\label{review}
In the introduction we have briefly discussed the necessity of filtering robust information in correlation based systems. The tools used to extract such an information can be divided essentially in three branches, specifically spectral methods, clustering methods and correlation based graphs.\\

Spectral methods are used to deal with multivariate systems since the beginning of 20th century. The most famous technique is the Principal Component Analysis (PCA) \cite{Mardia}. The idea behind PCA is to evaluate eigenvalues of the correlation matrix of the system, retain the $k$ greatest eigenvalues and project the empirical multivariate signal onto the sub-space generated by the corresponding $k$  eigenvectors. This approach allows one to reduce the complexity of the system. Indeed by indicating with $n$ the dimensionality of the system, usually $k\ll n$. At the same time, taking the first $k$ eigenvalues guarantees that the percentage of variance explained by the $k$ eigenvectors is the maximum value allowed by a $k$ mode projection of the data set \cite{Mardia}. Finally, a factor model \cite{Mardia} explaining the filtered correlation matrix can be constructed. The factor model allows one to simulate multivariate time series of arbitrary length $T$ such that, when T approaches infinity, the correlation matrix of simulated time series approaches exactly the correlation matrix filtered by PCA. 
A problematic aspect of PCA is that the number $k$ of retained eigenvalues is a free parameter of the method. 
The Random Matrix Theory gives the spectral distribution of the correlation matrix obtained from $n$ time series of length $T$ of independent Gaussian variables asymptotically  over $T$ and $n$ with fixed ratio $T/n > 1$ \cite{Marcenko67, Silverstein95, Sengupta99}. The Random Matrix Theory indicates the maximum eigenvalue $\lambda_{max}$ consistent with the hypothesis of independent random variables. This result can be applied to empirical correlation matrices to select the number $k$ of retained eigenvalues. All the eigenvalues greater than $\lambda_{max}$ should be retained \cite{Laloux99,Plerou99,Lillo05}. 
Such an approach has been fruitfully applied to the analysis of financial markets \cite{Laloux99,Plerou99}. The main findings relate to the interpretation of retained eigenvalues and corresponding eigenvectors. Usually the eigenvector associated to the greatest eigenvalue has components roughly degenerate, meaning that all stocks participate almost in the same way to the first eigenvalue. This fact suggested to interpret the first eigenvector as the so called ``market mode" \cite{Laloux99,Plerou99,Lillo05}. Other eigenvalues beside the greatest one resulted to be incompatible with the hypothesis of independent variables. Usually the corresponding eigenvectors have many near-zero components and then they should reveal communities of stocks in the market. Such communities involve stocks belonging to one or a few economic sectors \cite{Plerou02}.
%

Clustering techniques are used to reveal communities of elements in correlation based systems. The idea is that elements belonging to the same community share more information than elements belonging to different communities. The shared information is evaluated in terms of a similarity measure such as, for example, the correlation coefficient. The aim of clustering is to group together similar elements to get a significant partition of the system into communities (clusters). For a review of the most used techniques see for instance ref. \cite{Anderberg}. Newer techniques are for instance the superparamegnetic clustering of data \cite{Domany96}, which has been applied to financial data in Ref. \cite{Kullmann00}, the maximum likelihood data clustering \cite{Marsili02b}, the sorting points into neighborhoods \cite{Domany05}. Clustering procedures can also be used to identify a skeleton of interactions in the system. This is provided for instance by hierarchical clustering procedures in which communities are overlapping. Specifically, they are organized in a nested hierarchical structure \cite{Anderberg,Simon}. The idea of hierarchical clustering is that elements are sharing information according to the communities they belong to and communities are organized in a nested structure. We shall clarify this point with a simple example of nested communities. Consider two very important corporations like Microsoft Corp. and Intel Corp.. According to a standard economic classification both stocks belong to the economic sector of Technology, i.e. they belong to the same community at the level of economic sector. Further specification of the activity of stocks is possible. Microsoft Corp. is belonging to the economic sub-sector of  ``Software \& Programming" whereas Intel Corp. is belonging to the economic sub-sector of ``Semiconductors". This means that at the hierarchical level of sub-sectors Microsoft Corp. and Intel Corp. belong to different communities. \\
In this paper we shall concentrate on two hierarchical clustering techniques, i.e. the SLCA and the ALCA. The SLCA has been widely used to analyze the correlation between stocks in financial markets since the study of ref. \cite{Mantegna99}. Such an approach revealed that stocks belonging to specific economic sectors (e.g. the sector of Energy) cluster together at a high correlation level whereas stocks belonging to sectors such as Conglomerates never identify a community. Intermediate behavior has also been observed \cite{Mantegna99, Coronnello05}. The ALCA is widely used in the analysis of microarray data \cite{Eisen98}. In the study of multivariate biological systems this clustering procedure is usually preferred to other classical techniques of hierarchical clustering because of its property of averaging interactions. It has also been shown in ref. \cite{Coronnello05} that a comparative usage of SLCA, ALCA and Random Matrix Theory is suitable for the investigation of financial markets.\\

Graph theory is now widely used in the analysis of complex systems. Many physical, biological and social systems are suitably described by networks. Examples are Internet  \cite{Faloutsos99, Barabasi99, ALBarabasi02, Caldarelli2000, Pastor2001}, social networks \cite{Wasserman,Newman2002}, food webs \cite{Garlaschelli}, scientific citations \cite{Redner98} and sexual contacts among individuals \cite{Liljeros}.\\
Graph theory is also useful to extract information in correlation based systems. Consider a system of $n$ elements. The pair correlation coefficient between elements can be interpreted as the strength of the link connecting pairs of elements. In other words the system can be described by a complete network, i.e. a network of elements all connected one with each other, with weights associated to links according to the correlation coefficients. The topological structure of such a network is so simple (complete graph) that no information can be directly extracted from the topology. The idea is then to extract a subgraph from the complete network accounting for the weights of links. This procedure aims to translate the information contained in the correlation matrix of the system into the topological structure of the subgraph and then exploiting techniques of graph theory \cite{West, Newman03, Barabasi02} to analyze such an information.\\ 
A widely used subgraph of the complete network is the MST which is the spanning tree of shortest length \cite{West}. A discussion about the properties of the MST and an illustrative algorithm for its construction is given in section \ref{methods}. It is known \cite{Gower69} that the MST is deeply related to the SLCA (see next section). The number of links retained in the MST is $n-1$ for a system of $n$ elements and the tree is a connected graph. It has been shown that the MST is a powerful method to investigate financial systems \cite{Mantegna99}. In ref. \cite{Bonanno01} the MST has been used to analyze the system of the 100 most capitalized stocks traded in US equity markets. The aim of such a study was to understand how the structure of the correlation matrix changes when stock returns are evaluated at different time horizons. The result is that at short time horizons (5 min) the MST has a star like shape, the center of the star being the most capitalized stock of the system, i.e General Electric. 
When the time horizon is increased the star-like shape progressively disappears and branches of stocks mostly corresponding to specific economic sectors appear. These results suggest that at short time horizons the system is dominated by the ``market mode", whereas at longer time horizons the information about economic sectors becomes relevant. Another interesting result obtained by investigating the MST properties is related to the behavior of the market in proximity of a crash, such as the Black Monday, where a topological change of the tree is observed \cite{Onnela02,Onnela03}. \\
A generalization of the MST construction procedure relaxing the topological constraint that the resulting graph is a tree has been proposed in ref. \cite{Tumminello05}. The first level of generalization requires that the resulting graph is topologically planar, i.e. the graph can be drawn on a plane without crossing of links. Such a graph, named Planar Maximally Filtered Graph, allows topological structures forbidden in the MST such as loops and cliques \cite{West} that can be relevant for the analysis of correlation based systems. Specifically, the introduction of the ``connection strength" \cite{Tumminello05, Vespignani04} allows one to estimate the strength of connections of elements belonging to the same community and to investigate the structure of connections between communities such as, for example, stocks belonging to the same economic sector \cite{Tumminello05,Coronnello05}.

%
Another way to construct a subgraph from the complete network is to introduce a correlation threshold and then to remove those links with a correlation coefficient smaller than the threshold. Despite of its simplicity this method has shown to be useful in the study of economic \cite{Onnela04} and biological \cite{Eguiluz05} systems. In ref. \cite{Onnela04} authors study how graph changes when one reduces the value of the correlation threshold. The results of such an analysis are compared with those obtained for random graphs. One of the main results is that the formation of connected components in the empirical graph is incompatible with the corresponding formation in random graphs. A characteristic of this approach is that it is highly improbable to obtain a filtered network connecting all elements via some path by retaining a number of links of the same order of the number $n$ of elements. Then with a number of links of order $n$ it is rather difficult to describe interactions between any pair of elements of the system because the resulting graph is disconnected.\\

\section{New methods in correlation based networks} \label{methods}
In subsection \ref{ALMSTmet} we outline the algorithm producing the ALMST whereas in subsection \ref{bootstrap} we describe the measure of link reliability obtained by exploiting bootstrap replicas of data.

\subsection{Average Linkage Minimum Spanning Tree}\label{ALMSTmet}
In order to show how it is possible to associate a spanning tree to the ALCA it is useful to consider first an algorithm that generates the MST and performs at the same time the SLCA of the system. Consider a system of $n$ elements with estimated correlation matrix ${\bf C}$ of elements $\rho_{ij}$. To describe the algorithm it is necessary to recall the concept of connected component of a graph $g$ containing a given vertex $i$. The connected component of $g$ containing the vertex $i$ is the maximal set of vertices $S_i$ (with $i$ included) such that there exists a path in $g$ between all pairs of vertices belonging to $S_i$. When the element $i$ has no links to other vertices then $S_i$ reduces just to the element $i$. 
The starting point of the procedure is an empty graph $g$ with $n$ vertices. The algorithm can be summarized in 6 steps: 

(i) Set ${\bf Q}$ as the matrix of elements $q_{ij}$ such that ${\bf Q}={\bf C}$. 

(ii) Select the maximum correlation $q_{hk}$ between elements belonging to different connected components  $S_h$ and $S_k$ in $g$ \footnote{At the first step of the algorithm connected components in $g$ are coinciding with single vertices.}. 

(iii) Find elements $u$, $p$ such that $\rho_{up}=\text{{\bf Max}}\left \{\rho_{ij}, \forall i \in S_h \text{ and } \forall j \in S_k \right \}$

(iv) Add to $g$ the link between elements $u$ and $p$ with weight $\rho_{up}$. Once the link is added to $g$, $u$ and $p$ will belong to the same connected component $S=S_h\bigcup S_k$. 

(v) Redefine the matrix ${\bf Q}$: 

\begin{equation}\label{singleadjust}
\left\{
\begin{aligned}
q_{ij} & =q_{hk}, \text{ if } i\in S_h \text{ and } j\in S_k \\
q_{ij} & =\text{{\bf Max}} \left\{ q_{pt}, p\in S \text{ and } t\in S_j, \text{ with } S_j\neq S \right \}, \text{ if } i\in S \text{ and } j\in S_j \\
q_{ij} & =q_{ij}, \text{ otherwise}; \\
\end{aligned}
\right.
\end{equation} 
%

(vi) If $g$ is still a disconnected graph then go to step (ii) else stop.\\
The resulting graph $g$ is the MST of the system and the matrix ${\bf Q}$ results to be the correlation matrix associated to the SLCA. Specifically the matrix ${\bf D}^<$ of elements $d^<_{ij}=\sqrt{2 \left (1- q_{ij}\right )}$ is the subdominant ultrametric distance matrix of the system \cite{Rammal86,MantegnaBook}. The presented algorithm is not the most popular or the simplest algorithm for the construction of the MST but it clearly reveals the relation between SLCA and MST. Indeed connected components progressively merging together during the construction of $g$ are nothing else but clusters progressively merging together in the SLCA.\\

By replacing eq. (\ref{singleadjust}) with

\begin{equation}\label{averageadjust}
\left\{
\begin{aligned}
& q_{ij}=q_{hk}, \text{ if } i\in S_h \text{ and } j\in S_k \\
& q_{ij}=\text{{\bf Mean}}  \left\{ q_{pt}, p\in S \text{ and } t\in S_j, \text{ with } S_j\neq S \right \}, \text{ if } i\in S \text{ and } j\in S_j \\
& q_{ij}=q_{ij}, \text{ otherwise};
\end{aligned}
\right.
\end{equation} 
 in the step (v) of  the above procedure one obtains an algorithm performing the ALCA, being ${\bf Q}$ the correspondent correlation matrix. The resulting tree $g$ that we call ALMST is the tree associated to such a clustering procedure. \\
The choice of the link at step (iii) of the ALMST construction algorithm does not affect the clustering procedure. More precisely by selecting any link between nodes $u \in S_h \text{ and } p \in S_k $ the matrix ${\bf Q}$ representing the result of ALCA remains the same. This degeneracy allows one to consider different rules to select the link between elements $u$ and $p$ at the step (iii) of the construction protocol. Different rules at step (iii) give rise to different correlation based trees. The same observation holds true for the algorithm that generates the MST. This fact implies that in principle one can consider spanning trees which are different from the MST and are still associated with the SLCA. However, it is worth recalling that the MST is unique in the sense that it is the spanning tree of shortest length \cite{West}. 
%

\subsection{The bootstrap value as a measure of link reliability}\label{bootstrap}

The bootstrap technique invented by Efron (1979) \cite{Efron79} is widely used in phylogenetic analysis since the paper by Felsestein (1985) \cite{Felsenstein85} as a phylogenetic hierarchical tree evaluation method (see for instance \cite{Efron96}). The basic idea behind the technique is simple. Consider a system of $n$ elements and suppose to collect data in a matrix ${\bf X}$ with $n$ columns and $T$ rows. $T$ is the number of collected samples, e.g. the length of time series. The process is assumed to be stationary. The correlation matrix ${\bf C}$ of the system is estimated from the matrix ${\bf X}$ by evaluating the pair correlation coefficient. By applying the procedures described in the previous subsection to ${\bf C}$ one can construct the MST and ALMST of the system. The bootstrap technique requires to construct a number $r$ of replicas ${\bf X}^*_i$, $i\in{1,...,r}$ of the data matrix ${\bf X}$. Usually in phylogenetic analysis $r=1000$ is considered a sufficient number of replicas. Each replica ${\bf X}^*_i$ is constructed by randomly selecting $T$ rows from the matrix ${\bf X}$ allowing repetitions. This procedure implies that some rows of the original data matrix are included in the replica more then once whereas other rows are missed in the replica. For each replica ${\bf X}^*_i$ the correlation matrix is evaluated and the MST and ALMST are extracted. The result is a collection of MSTs $\{MST^*_1,...,MST^*_{r} \}$ and ALMSTs $\{ALMST^*_1,...,ALMST^*_{r} \}$. To associate the so called \emph{bootstrap value} to a link of the original MST (ALMST) one evaluates the number of $MST^*_i$ ($ALMST^*_i$) in which the link is appearing and normalizes such a number with the total number of replicas, e.g. $r=1000$. The bootstrap value gives information about the reliability of each link of a graph.\\
It is to note first that the bootstrap approach does not require the knowledge of the data distribution and then it is particularly useful to deal with high dimensional systems where it is difficult to infer the joint probability distribution from data. Second, the average of bootstrap values in a graph can be considered as a global measure of the reliability of the graph itself and then can be used to compare different filtered graphs. In case of normally distributed random variables the error associated to the correlation coefficient $\rho$ roughly scales like $(1-\rho^2)/\sqrt{T}$. One might then be tempted to expect that the higher is the correlation associated to a link in a correlation based network the higher is the reliability of the link. We shall show in the following section that such a conjecture cannot explain empirical results for the system of stock returns we consider. Finally the graph weighted with bootstrap values can be helpful in the search for significant communities in the system.

\section{Empirical Analysis} \label{empirical}
We perform an empirical investigation by considering the 300 most capitalized stocks traded at NYSE during the time period 2001-2003. We consider the capitalization of stocks at December 2003. The return time series are sampled at daily time horizon by computing the logarithm of the ratio of closure to open price and the series length is $T=748$. Stocks are classified in terms of the economic sector and sub-sector they belong to, according to the classification scheme used in the website {\tt http://finance.yahoo.com}. Economic sub-sectors are giving a further specification of the activity of firms belonging to the same economic sector. For instance, stocks belonging to the sector of Transportation are partitioned in stocks belonging to the Railways sub-sector and in stocks belonging to the economic sub-sector of Airlines. The total number of economic sectors involved in our set of stocks is 12 whereas the number of sub-sectors is 78. The list of the 300 stocks considered with their economic sector and sub-sector is available at  {\tt http://lagash.dft.unipa.it/IJBC.html}.

\begin{widetext}

\begin{figure}
\includegraphics[width=0.95\textwidth]{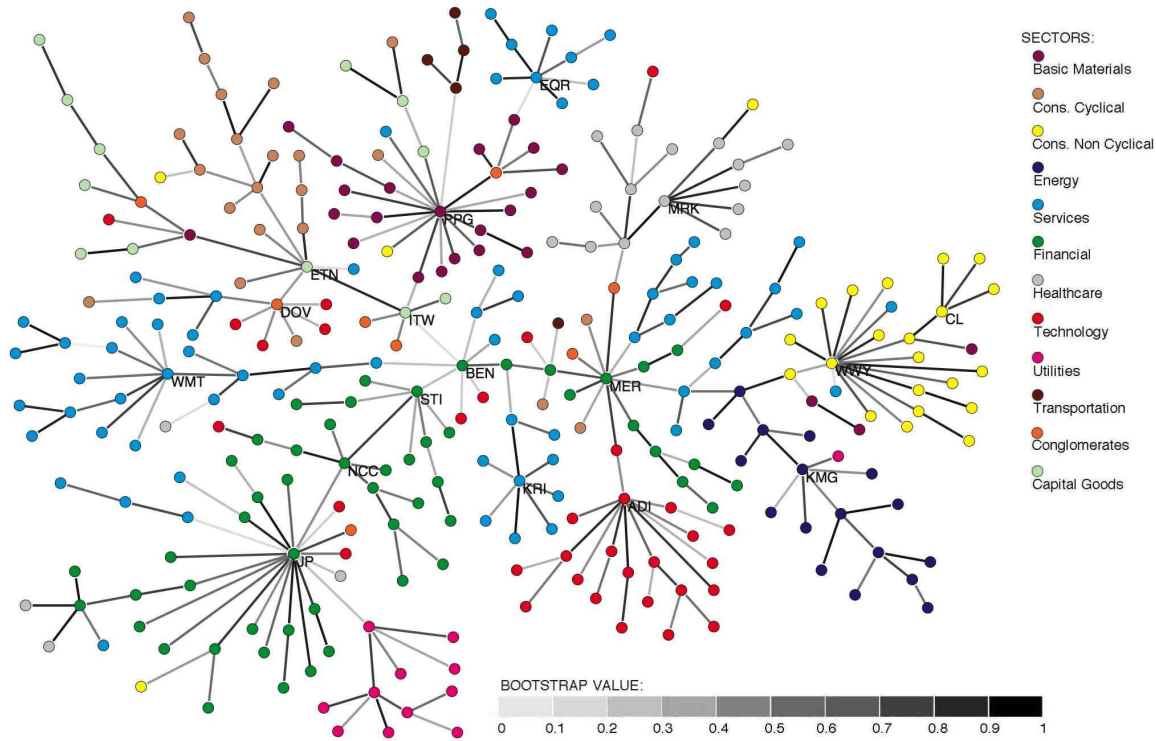}
\caption{Minimum Spanning Tree associated to daily price returns of the 300 most capitalized stocks belonging to NYSE traded in the time period 2001-2003. Economic sectors are highlighted using different vertex colors (see text for details). Links are drawn using a grey scale with 10 levels. Darker links have a higher bootstrap value. A version of this figure with the tick symbol associated to each stock and the list of links of the MST with the corresponding values of correlation and bootstrap are available at  {\tt http://lagash.dft.unipa.it/IJBC.html}.} 
\label{MST}
\end{figure}

\end{widetext}

\begin{widetext}

\begin{figure}
\includegraphics[width=0.95\textwidth]{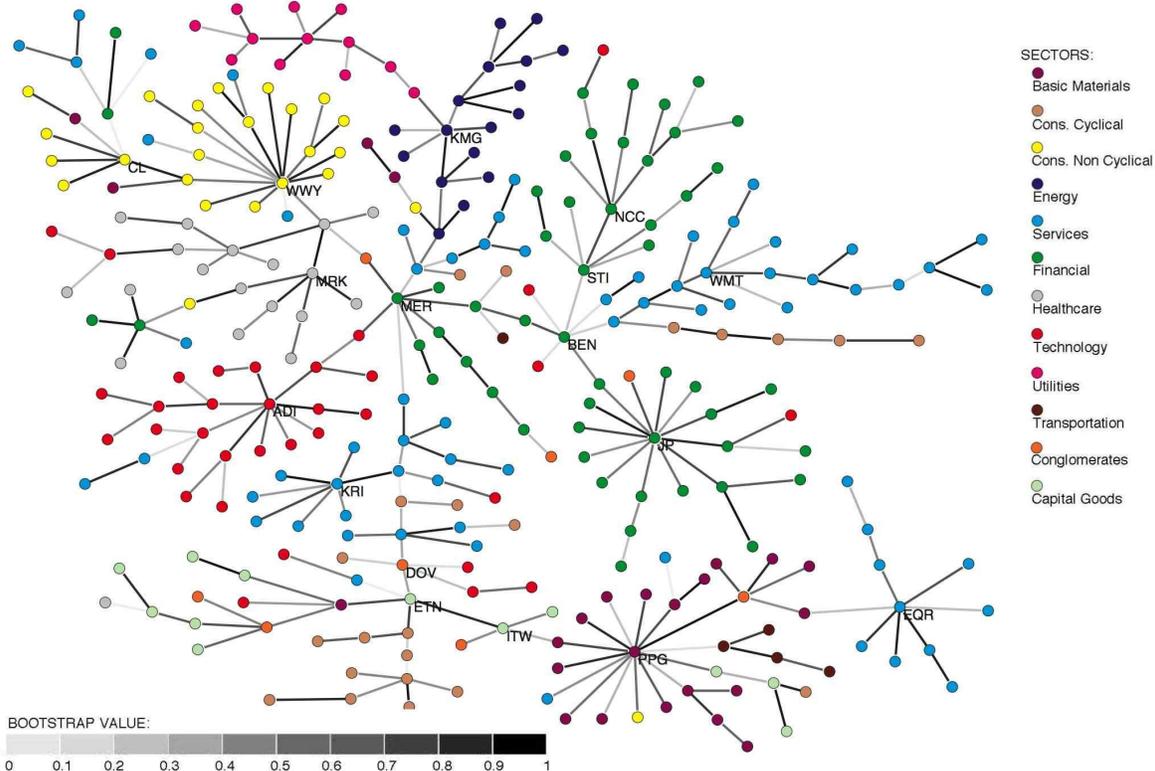}
\caption{Average Linkage Minimum Spanning Tree associated to daily price returns of the 300 most capitalized stocks belonging to NYSE. Transactions occurred in 2001-2003 are considered. Economic sectors are highlighted using different vertex colors (see text for details). Links are drawn using a grey scale with 10 levels. Darker links have a higher bootstrap value. A version of this figure with the tick symbol associated to each stock and the list of links of the ALMST with the corresponding values of correlation and bootstrap are available at  {\tt http://lagash.dft.unipa.it/IJBC.html}.} 
\label{ALMST}
\end{figure}

\end{widetext}

In Figs. \ref{MST} and \ref{ALMST} the MST and ALMST of the system are shown respectively. Links are drawn using a grey scale with 10 levels selected according to the bootstrap value. The bootstrap value is evaluated over $r=1000$ replicas. The higher is the bootstrap value associated to a link the darker is the line representing the link. Vertices are drawn with different colors to highlight the economic sector each stock belongs to. Specifically these sectors are Basic Materials (violet, 24 stocks), Consumer Cyclical (tan, 22 stocks), Consumer Non Cyclical (yellow, 25 stocks), Energy (blue, 17 stocks), Services (cyan, 69 stocks), Financial (green, 53 stocks), Healthcare (gray, 19 stocks), Technology (red, 34 stocks), Utilities (magenta, 12 stocks), Transportation (brown, 5 stocks), Conglomerates (orange, 8 stocks) and Capital Goods (light green, 12 stocks). For the sake of readability of the pictures, the tick symbol is reported only for stocks with degree greater than 5 in at least one of the graphs.\\
A comparison of the graphs shows a number of similarities. Indeed the $85\%$ of links are common to both the MST and ALMST. In fact the intra-sector structures are similar in the MST and ALMST. Moreover the degree of highly connected stocks (degree greater than 5) is similar in both graphs. There are also differences between the graphs. 
These differences are mostly observed in the connections between groups of stocks of different economic sectors. For instance the Consumer Non Cyclical economic sector shows a star like shape with WWY as hub in both graphs but the path connecting this sector to others is different. In the MST the sector of Consumer Non Cyclical is connected via a high reliable link to the sector of Energy and then, following a path traversing the sectors of Services, Financial and Conglomerates, it is connected to the Healthcare sector. Differently, in the ALMST the sector of Consumer Non Cyclical is directly connected to the Healthcare sector and far from the Energy sector. Also the position of the sector of Basic Materials is different in the two graphs. Specifically in the MST  this sector is two steps far from the Financial sector whereas in the ALMST a path as long as 9 steps is necessary to connect stocks of Basic Materials to the Financial sector. \\

%


\begin{figure}
\includegraphics[width=0.5\textwidth]{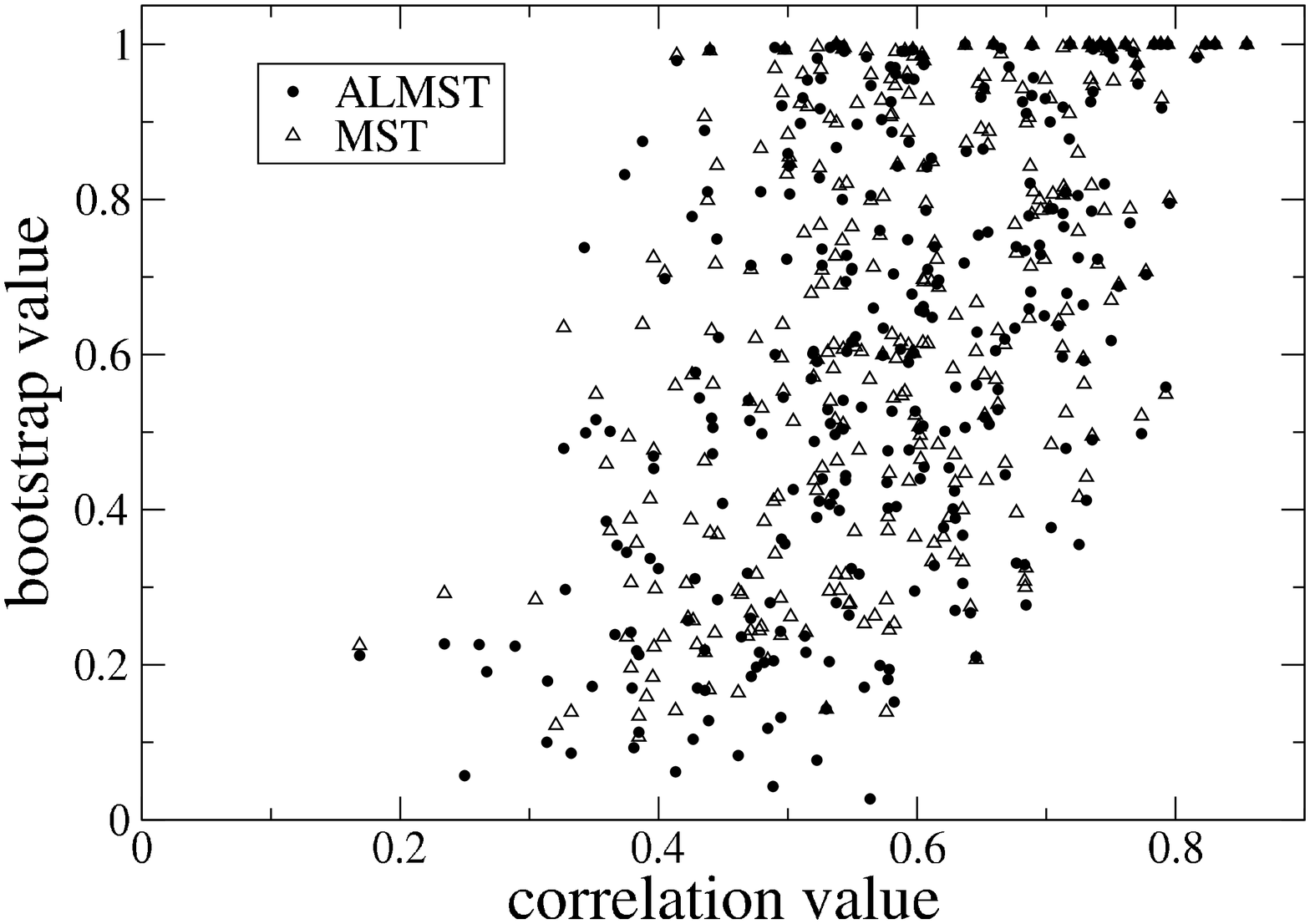}
\caption{Scatter plot of the bootstrap value and cross correlation of links in ALMST (black circles) and in MST (open triangles)} 
\label{bootvsCCmax}
\end{figure}

In Fig. \ref{bootvsCCmax} we report the scatter plot of the bootstrap value against the correlation coefficient associated to links of both the ALMST and MST. Black circles refer to the ALMST and open triangles refer to the MST. The figure clearly shows that the reliability of links cannot be straightforwardly explained in terms of the correlation coefficient value associated to links. In fact a large dispersion of points is evident, although we observe a significant positive correlation between the link bootstrap reliability and the correlation coefficient value. Indeed we observe links associated with a quite small correlation, e.g. 0.4, having a very high reliability in terms of the bootstrap value, e.g. 1. On the contrary some links with correlation as high as 0.7 show a small reliability of only 0.3 in terms of the bootstrap value. This result suggests that the strength of a link and its reliability in correlation based networks are carrying partially different information.
%

In order to assess the ability of MST and ALMST in detecting communities defined in terms of economic sectors and sub-sectors, we have counted the number of intra-sector links and intra-subsector links in both the graphs. In the MST 223 links connect stocks  of the same economic sector and 139 links are intra-sub-sector links. In the ALMST we have counted 227 intra-sector links and 144 intra-sub-sector links. These results suggest that ALMST detects communities defined in terms of economic sectors and sub-sectors slightly better than the MST, for the specific system we are dealing with. 
\begin{table}
\label{averBoottab}
\caption{bootstrap value averages ($<bv>\pm \sigma_{<bv>}$)}
\begin{tabular}{||c|c|c||}
\tableline
 considered links& $ALMST$ & $MST$\\
\tableline
\tableline
all links & $0.602\pm0.016$ & $0.627\pm0.015$ \\
intra-sector & $0.657\pm0.017$ & $0.680\pm0.016$ \\
inter-sector & $0.426\pm0.035$ & $0.469\pm0.030$ \\
intra-sub-sector & $0.725\pm0.020$ & $0.740\pm0.019$ \\
inter-sub-sector & $0.488\pm0.022$ & $0.529\pm0.020$ \\
inter-sub-sec. \& intra-sec. & $0.540\pm0.026$ & $0.582\pm0.026$ \\
\tableline
\tableline
\end{tabular}
\end{table}

In Table I we report results obtained by averaging the bootstrap value of links grouped in different classes for both MST and ALMST. By comparing the column of values corresponding to the MST with the one corresponding to the ALMST we note that links belonging to the MST are in average more reliable than links belonging to the ALMST. We also observe that the average bootstrap value of links connecting stocks belonging to the same economic sub-sector is greater than $0.7$ in both the graphs and it is immediately followed by the average bootstrap value of intra-sector links, which is 0.657 in the ALMST and 0.680 in the MST. Both the average bootstrap values of intra-sector links and intra-sub-sector links are greater than the average bootstrap value evaluated over all links in the graphs. This evidence suggests that sectors and sub-sectors are significant communities in both networks. Such an indication is also supported by simulations. Indeed the stronger reliability of intra-sector links and intra-sub-sector links with respect to inter-sector links and inter-sub-sector links has been also observed in data simulations based on a 3-level Hierarchically Nested Factor Model \cite{TumminelloHope06}. In this model each stock is depending on 3 factors, the first one being associated to the economic sub-sector, the second one being relative to the economic sector and the last one to the market. We shall discuss this investigation in a forthcoming paper.\\ 

\section{Conclusions}\label{conclusions}
We have introduced a technique allowing to extract a correlation based tree named ALMST associated with the ALCA. We have also introduced a new measure of the reliability of links based on the bootstrap technique. We have applied both techniques to the system of daily returns of the 300 most capitalized stocks traded at NYSE. For this system a comparison with the MST indicates a slightly greater capability of the ALMST in recognizing  economic sectors and sub-sectors in market return data, whereas links of the MST are in average more reliable in terms of the bootstrap value than links belonging to the ALMST. 
We have also shown that the reliability of links cannot be explained just in terms of the strength of connections and that intra-sector and intra-sub-sector connections are in average more reliable, with respect to the bootstrap technique, than inter-sector and inter-sub-sector links in both MST and ALMST. Such a result suggests to interpret economic sectors and sub-sectors as significant communities in the market. 

\section{Acknowledgments}
Authors acknowledge support from the research project MIUR 449/97 ``High frequency dynamics in financial markets", the research project MIUR-FIRB RBNE01CW3M ``Cellular Self-Organizing nets and chaotic nonlinear dynamics to model and control complex systems" and from the European Union STREP project n. 012911 ``Human behavior through dynamics of complex social networks: an interdisciplinary approach".



\end{document}